\documentclass[aps,pra,twocolumn,superscriptaddress,10pt]{revtex4-1}

\usepackage[english]{babel}
\usepackage{amsmath}
\usepackage{amsthm}
\usepackage{amssymb}
\usepackage{mathrsfs}
\usepackage{booktabs}
\usepackage{color}
\usepackage{hyperref}
\usepackage[ruled,vlined]{algorithm2e}
\usepackage{cases}
\usepackage{enumitem}

\usepackage{multirow}
\hypersetup{
  colorlinks   = true,  
  urlcolor     = blue,  
  linkcolor    = blue,  
  citecolor    = red    
}
\usepackage{mathtools}
\usepackage{natbib}
\usepackage{pgfplots}
\pgfplotsset{compat=1.18}
\usepackage{bbm}
\usepackage{subfigure}
\usepackage{url}
\usepackage{tabularx}
\newcolumntype{P}[1]{>{\centering\arraybackslash}p{#1}}

\newcommand{\di}{{\rm d}}

\mathtoolsset{showonlyrefs}

\theoremstyle{definition}

\theoremstyle{remark}

\theoremstyle{plain}
\newtheorem*{theorem*}{Theorem}

\usepackage{scalerel,stackengine}
\stackMath
\newcommand\reallywidehat[1]{%
\savestack{\tmpbox}{\stretchto{%
  \scaleto{%
    \scalerel*[\wi\di thof{\ensuremath{#1}}]{\kern-.6pt\bigwedge\kern-.6pt}%
    {\rule[-\textheight/2]{1ex}{\textheight}}
  }{\textheight}%
}{0.5ex}}%
\stackon[1pt]{#1}{\tmpbox}%
}
\usepackage{comment}
\usepackage{float}

\SetArgSty{textnormal}

\makeatletter

\newenvironment{widetext2}{%
  \par\ignorespaces
  \setbox\widetext@top\vbox{%
   \vskip15\p@
   \hb@xt@\hsize{%
    \leaders\hrule\hfil
    \vrule\@height6\p@
   }%
   \vskip6\p@
  }%
  \setbox\widetext@bot\hb@xt@\hsize{%
    \vrule\@depth6\p@
    \leaders\hrule\hfil
  }%
  \onecolumngrid
  \let\set@footnotewidth\set@footnotewidth@ii
}{%
  \par
  \twocolumngrid\global\@ignoretrue
  \@endpetrue
}%

\makeatother

\begin{document}

\title{Pulse family optimization for parametrized quantum gates using spectral clustering}

\author{R.J.P.T. \surname{de Keijzer}}
\affiliation{Department of Applied Physics, Eindhoven University of Technology, P. O. Box 513, 5600 MB Eindhoven, The Netherlands}
\affiliation{Eindhoven Hendrik Casimir Institute, Eindhoven University of Technology, P. O. Box 513, 5600 MB Eindhoven, The Netherlands}
\altaffiliation[Corresponding author: ]{r.j.p.t.d.keijzer@tue.nl }

\author{J.A.C. \surname{Snijders}}
\affiliation{Eindhoven Hendrik Casimir Institute, Eindhoven University of Technology, P. O. Box 513, 5600 MB Eindhoven, The Netherlands}
\affiliation{Q-CTRL, Berlin, Germany}
\author{A. R. R. \surname{Carvalho}}
\affiliation{Q-CTRL, Berlin, Germany}

\author{S.J.J.M.F. \surname{Kokkelmans}}
\affiliation{Department of Applied Physics, Eindhoven University of Technology, P. O. Box 513, 5600 MB Eindhoven, The Netherlands}
\affiliation{Eindhoven Hendrik Casimir Institute, Eindhoven University of Technology, P. O. Box 513, 5600 MB Eindhoven, The Netherlands}

\date{\today}

\begin{abstract}
Parametrized gate circuits are used in plentiful applications in the current NISQ era of quantum computing. These parametrized gates are chiefly implemented using analytically found pulse protocols, often yielding suboptimal gate times, and consequently, fidelities. Alternatively, gate optimization algorithms are designed to construct high fidelity pulses for individual, fixed points in continuous parameter space. Gates for intermediate parameters can subsequently be found by some form of interpolation between previously constructed pulses. Nevertheless, it is not guaranteed (as with analytic protocols) that the pulses found by the optimization algorithms belong to the same \textit{family} of solutions and thus show resemblance. Interpolation between two pulses of differing solution families often leads to high infidelities, as the pulse strays away from the minimum in the parameter/fidelity landscape. In this work, we introduce a \textit{spectral clustering} method to sort high-fidelity, optimized pulses in families, and interpolating solely between pulses of the same family. Accordingly, interpolations will always approach maximal fidelity. Furthermore, as more than one pulse family is constructed, the parameter space can be partitioned according to which family prevails fidelity-wise. This work provides a meticulous demonstration of our constitutive continuous gate family construction by applying it to a universal gate set for Rydberg and Cat qubits under noise.
\end{abstract}

\maketitle

\section*{INTRODUCTION}
\label{sec:introduction}

In order for a quantum computer to be able to perform all possible computations, it has to possess the ability to execute a universal gate set, consisting of all possible single qubit rotations and at least one entangling two-qubit operation \cite{universality}. The canonical single qubit gate set are the rotations gates over angle $\theta$: $R_X(\theta), R_Y(\theta),$ and $R_Z(\theta)$, which form the basis for many variational quantum algorithms (VQAs) \cite{keijzer1}. These parametrized gates are executed in physical systems using some control function on the qubits, i.e. laser pulses or electrical currents, specific to each individual parameter. Current era quantum computing systems are in the NISQ-era, where qubits are highly susceptible to noise \cite{nisq}. This means that it is important to choose the optimal control function to mitigate noise, which is often not a standard analytical protocol, but some faster control pulse as constructed by a pulse optimization algorithm \cite{madhav,sven,boulderopal1,knill,crab,vqoc}. These variational methods often lead to faster, and consequently higher fidelity, pulses.

The problem with these optimization algorithms is that, unlike analytical protocols, they only prescribe pulses for one fixed point in the parameter space at a time, and not for the entire continuous set of parameters, which is especially relevant for VQAs \cite{continuousgatesetimportant,Grange2023,Cerezo2021}, where copious different parameter realizations are necessary. Constructing control pulses every time a new unique parameter is required is far too computationally expensive, making it infeasible in the majority of VQAs.

\begin{figure}[H]
    \centering
\includegraphics[width=0.95\columnwidth]{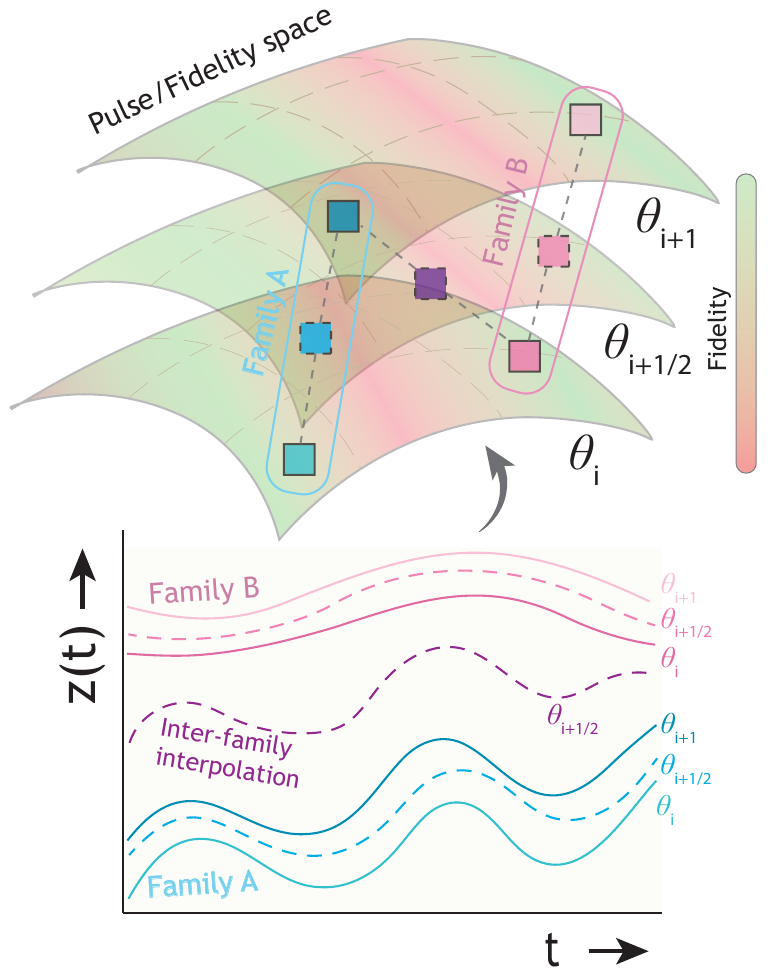}
    \caption{Graphical representation of the pulse interpolation problem. A pulse optimization algorithm might find one of several local minimizing pulses for $\theta_i$ and $\theta_{i+1}$. If these are corresponding local minima, the pulses belong to the same family (blue or pink) and interpolation yields a pulse for $\theta_{i+1/2}$ with high fidelity. If the minima do not correspond, interpolation happens between pulses of different families, yielding a pulse for $\theta_{i+1/2}$ with low fidelity (purple).}
    \label{fig:graphabstract}
\end{figure}

The task at hand is to extend a collection of solutions for discrete points in parameter space to the entire continuous set by some form of interpolation. However, this approach has an underlying complication. Assume a set of $M$ subsequent, close points in the gate parameter space $\{\theta_i\}_i^M$ for which one wishes to optimize the pulses. Initially, the pulse optimization algorithm will find a local minimizing pulse for $\theta_0$, and because the optimization problem is regular, there will exist a corresponding local minimizing pulse for $\theta_{1}$ as long as the two values for $\theta$ are adequately close. In general, there will exist corresponding pulse solutions for each $\theta_i$, which we then call a pulse \textit{family}, as introduced in Ref.~\cite{families3}. Interpolating within a pulse family will yield pulses for intermediate values of $\theta$, with comparable high fidelities. The problem, however, is that there is no guarantee that for every value of $\theta_i$, matching local minima will be found, and thus there is no guarantee that all the optimized pulses found will belong to a single family. Interpolating \textit{between} two pulse families, will generally yield lower fidelities because the resulting pulses will not be in local minima of the pulse/fidelity landscape, see Fig.~\ref{fig:graphabstract}.

Previous work has been done on the construction of high fidelity continuous gate sets. References~\cite{families4,madhav,vqoc} construct pulses to mitigate hardware errors based on a parameterized Hamiltonian. These results are extended to the continuous case in~\cite{optimalpulse1}. Reference~\cite{cubicinteraction} is the first to mention the problem of similarity between pulses, but only related to single parameter optimization. References~\cite{families3,families2,families5,families1} all devise methods to optimize continuous gate sets while maintaining pulse similarity across different parameters for interpolation purposes. In~\cite{families3,families2}, a neural network approach is used, where the networks are trained by randomly sampling the parameter space. The disadvantage of such a black box method is that verification of the resulting pulses is non-trivial. In~\cite{families5}, Trotterization is used to implement continuous parameter exponential matrices, but does not account for time optimization. The work in~\cite{families1} utilizes Tikhonov regularization and feedforward to enforce pulse similarity, which is partly employed in our method. However, our work is the first to utilize clustering methods from graph theory \cite{graphcluster1,graphcluster2} to find multiple families of gates, which has three main advantages over all previous methods:
\begin{itemize}[leftmargin=1.5em,itemsep=0pt]
    \item[--] Foremost, partitioning the parameter space according to where certain pulse families outperform the others has the potential to yield exceedingly higher fidelities than a singular family approach;
    \item[--] Finding multiple families allows for the selection of pulses which are best suited to experimental procedures after verification on a setup;
    \item[--] The clustering method provides a more well-suited manner of interpolating pulses using Wasserstein-2 distances from optimal transport (see Sec.~\ref{fig:wasserstein}).
\end{itemize}

The layout of this paper is as follows. Section~\ref{sec:pulseoptimization} describes optimal control methods used to construct optimal pulses for discrete parameter values $\theta_i$. Section~\ref{sec:clustering} prescribes our pulse family clustering algorithm, its relation to other methods, as well as its advantages. Section~\ref{sec:pulseinterpolation} details the interpolation methods employed in this work. In Sec.~\ref{sec:results}, we show initial results for our approach, starting with results on the Wasserstein-2 distance to characterize pulse similarity in Sec.~\ref{sec:wasserstein}. Section~\ref{sec:interpolation} illustrates the increases in fidelities our method has for interpolation. Finally, Sec.~\ref{sec:conclusion} summarizes and presents a future outlook.

\section{Pulse Optimization}
\label{sec:pulseoptimization}
The application analyzed in this work is creating a universal parametrized gate set of $R_X(\theta), R_Y(\theta), R_Z(\theta)$ and $R_{ZZ}(\theta)$, where $\theta\in[0,\pi]$. This will be performed on Rydberg qubits \cite{rydberg1,rydbergqubit}, an architecture that recently has become exceedingly mature, and also on Cat qubits \cite{catqubitmain,kerr1}, which recently are gaining traction within the quantum computing community. This illustrates the versatility of our methods. In this section, we briefly describe the optimization procedure for fixed parameters $\theta_i$, which has been well-described in literature before \cite{keijzer1,pulseopt1,pulseopt2}. The evolution of the qubit system density matrix $\rho_t$ follows a Lindblad equation \cite{lindblad1} of the form
\begin{equation}
\label{eq:lindblad}
\begin{aligned}
    \partial_t\rho_t&=-i\big[H_{\text{sys}}[z(t)],\rho_t\big]+\mathcal{L}(\rho_t), \quad \rho(0)=\rho_0,\\
    \mathcal{L}(\rho)&=\sum_k \gamma_k V_k \rho V_k-\frac{1}{2}\gamma_k\{V_k^\dagger V_k,\rho\},
\end{aligned}
\end{equation}
where $H_{\text{sys}}[z(t)]$ is the system Hamiltonian controlled by user-defined, and optimizable pulses $z(t)$, and $\mathcal{L}$ is the Lindblad operator responsible for the different sources of decoherence in the system. $V_k$ and $\gamma_k$ respectively are the jump operators with corresponding strengths, defining the Lindblad operator. The pulses are functions on $[0,T]$, where $T$ is the gate end time, bounded by experimental limitations as $z_{\text{min}}\leq z(t) \leq z_{\text{max}}$. The physical interpretation of the pulse type is system specific. In Rydberg systems, we can control the transitions between the $|0\rangle$, $|1\rangle$ and $|r\rangle$ states on qubit $j$ using coupling strengths $\Omega_{ab,j}$ and detunings $\Delta_{b,j}$ on transition $|a\rangle\leftrightarrow|b\rangle$ to get $z(t)\in\{\Omega_{01,j}(t),\Delta_{1,j}(t),\Omega_{1r,j}(t),\Delta_{r,j}\}$ \cite{keijzer1}. Meanwhile, Cat qubits offer control on the single photon drive $E_j(t)$, the detuning $\Delta_j(t)$, interaction strength $g(t)$, and the two photon drive $G_j(t)$ \cite{catqubitmain}. This gives  $z(t)\in\{E_j(t),\Delta_{j}(t),G_{j}(t),g(t)\}$. For the Rydberg and the Cat qubit systems, we introduce characteristic timescales $\tau_{\text{Ryd}}=1/\Omega_{\text{max}}$ and $\tau_{\text{Cat}}=1/K$ respectively, where $\Omega_{\text{max}}$ is a maximal coupling strength in the Rydberg system and $K$ is the Kerr non-linearity in the Cat qubit system. Since our methods are agnostic to the qubit architecture and the physical implementation is not integral to understanding our results, we do not further discuss details on the specific systems here. For detailed Hamiltonians, Lindblad operators and a discussion on the pulse end times $T$, see App.~\ref{app:hamiltonians}.

A solution to the Lindblad equation~\eqref{eq:lindblad} for a specific set of pulses $z(t)$ is described by a trace preserving, completely positive operator called a quantum channel $\mathcal{E}_z$, s.t. $\rho_T=\mathcal{E}_z(\rho_0)$ \cite{quantumchannel}. For each specific gate $U\in\{R_X(\theta_i), R_Y(\theta_i), R_Z(\theta_i)$, $R_{ZZ}(\theta_i)\}$ to be created, the goal is to optimize the fidelity given by
\begin{equation}
\label{eq:cost}
    \min_z F(\mathcal{E}_z, \hat{U}):=\min_z\frac{\sum_j \operatorname{Tr}\left(U P_j^{\dagger} U^{\dagger} \mathcal{E}_z\left(P_j\right)\right)+d^2}{4^{N}d+d^2},
\end{equation}
as in \cite{fidelity2}. Here $N$ is the number of qubits and $d$ is the dimension of the qubit system. For Rydberg systems $d=3^{N}$ and for Cat systems $d=2^{N}$. $P_j$ are the Pauli matrices (for Rydberg extended to a 3-dimensional system) of which there are $4^{N}$. Using the software package Boulder Opal, we optimize the pulses $z(t)$ for the cost function in~\eqref{eq:cost}, using the package's standard convergence criteria \cite{boulderopal1, boulderopal2}. For all problems in this work, we optimize $M=20$ equidistant predefined angles $\{\theta_i\}_i^M$ in the parameter space $[0,\pi]$.

\section{Pulse clustering}
\label{sec:clustering}
\begin{figure}
    \centering
    \includegraphics[scale=0.50]{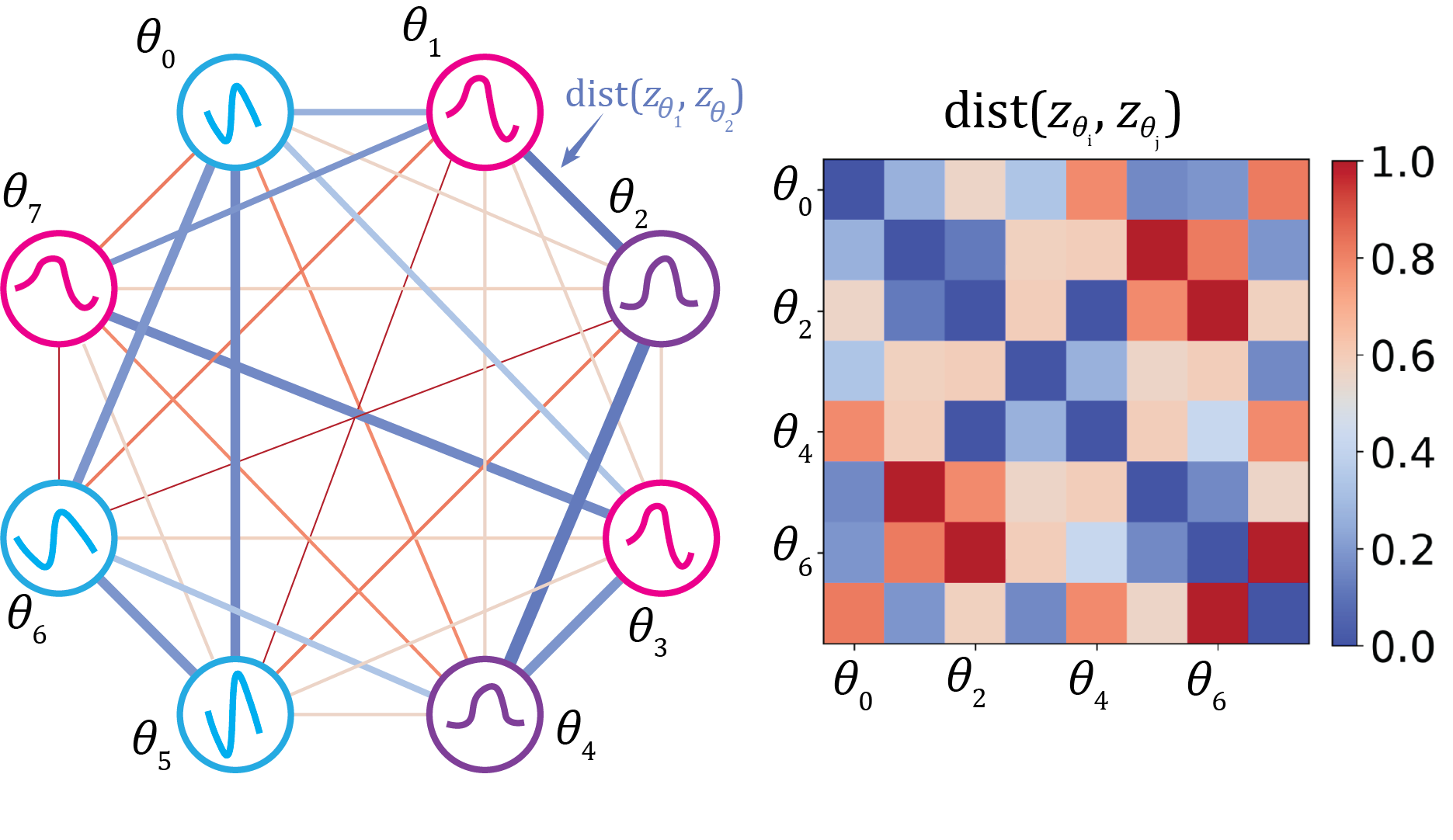}
    \caption{To distinguish between different pulse families, pulses are represented as nodes in a graph with the edge weighs between them equal to some distance function $d$. The spectral clustering algorithm takes as input the distance matrix (right), and creates clusters minimizing the distances between pulses of the same cluster.}
    \label{fig:pulsegraph}
\end{figure}

After having optimized the pulses for discrete fixed parameters $\{\theta_i\}_{i}^M$, they need to be partitioned in corresponding solution branches, or families. Here, we detail our method of clustering pulses in a fixed number of families using \textit{spectral clustering}~\cite{graphcluster2}. This method sorts the nodes $V$ of a weighted graph $G=(V,E,w)$ into a fixed number of clusters. $E$ are the edges of the graph, and $w$ are the weights on these edges encoding the similarity of two nodes. To apply this method to the discrete pulse sorting, we define a fully-connected graph, where each node $i$ corresponds to one of the optimized pulses found for a specific parameter $\theta_i$, as shown in Fig.~\ref{fig:pulsegraph}. The weights are set as $w(z_1,z_2)=(\text{dist}(z_1,z_2)+\epsilon)^{-1}$, where $\text{dist}$ is some distance on the space of pulses and $\epsilon=10^{-4}$ is a small regularization parameter. As a basic approach, the $L_2$ distance \cite{l2norm} can be used as
\begin{equation}
    L_2(z_1,z_2)=\int_{0}^T |z_1(t)-z_2(t)|^2dt.
\end{equation}

\begin{figure}
    \centering
    \includegraphics[scale=0.44]{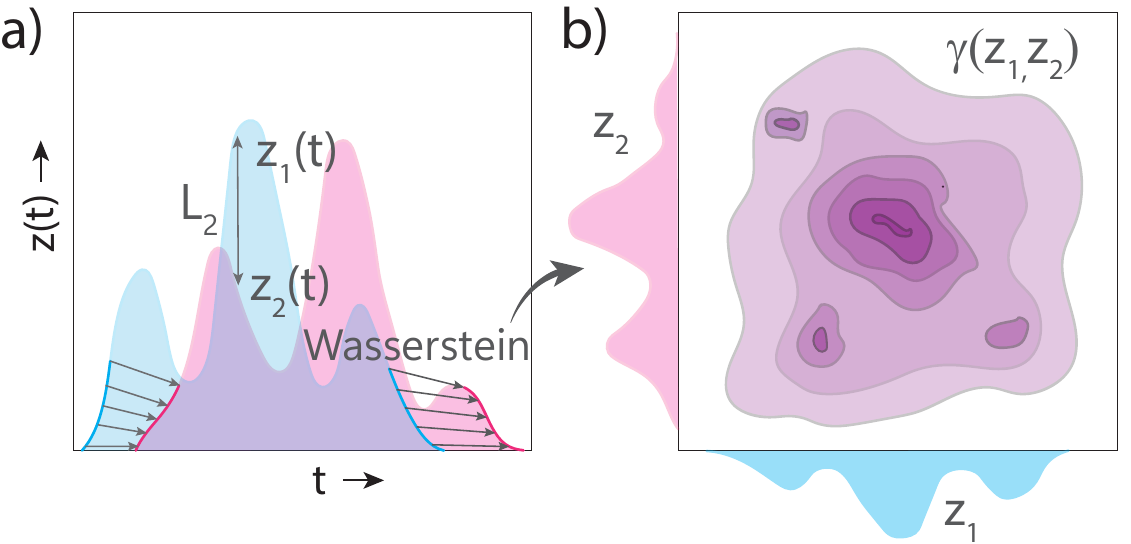}
    \caption{Example of Wasserstein-2 vs. $L_2$ distance. a) If two pulses are compared in $L_2$ distance, small or large time shifts both can result in maximal dissimilarity. The Wasserstein-2 distance is able to capture these shifts because it tries to transport one pulse onto the other. b) Example of a coupling $\gamma\in\Pi$ between pulses $z_1$ and $z_2$, the Wasserstein-2 distance aims to construct a minimal coupling in the sense of~\eqref{eq:wmetric}. }
    \label{fig:wasserstein}
\end{figure}

This distance function is often implemented as it is efficient to compute. However, it does not fully serve our needs, since it does not capture shifts on the time axis well. The $L_2$ distance only compares pulse values pointwise in time (see Fig.~\ref{fig:wasserstein}, where the highest peak shifts, causing a big $L_2$ distance between the pulses even though they are quite similar in shape). As long as two pulses do not overlap, a large time shift causes the same maximal $L_2$ distance as a small shift. To remedy this issue, the Wasserstein-2 distance~\cite{santambrogio} can be used. This distance has its origins in the mathematical framework of optimal transport and defines the cost of moving one distribution onto another, which is more closely related to our intuitive likeness of pulses, see Fig.~\ref{fig:wasserstein}. The important point to understand is that the Wasserstein-2 distance treats the pulses as distributions on the plane $\mathbb{R}^2$ and calculates the cost of moving one onto the other and defines this as the distance \cite{villani}. The exact definition of the distance is rather mathematically involved, and therefore deferred to App.~\ref{app:Wasserstein} \footnote{For the Wasserstein-2 distance, pulses are discretized as point clouds with points equidistant along the curve, and Sinkhorn-Knopp \cite{sinkhorn} is used to identify the minimizing coupling.}.

Using one of the previously defined distances, a similarity matrix consisting of the edge weights $w$ can be constructed, see Fig.~\ref{fig:pulsegraph}. This serves as the input for the spectral clustering algorithm. As the spectral clustering algorithm requires a fixed number of clusters, we determine the best possible number of clusters by means of the \textit{Elbow method} \cite{elbow}. This heuristic determines the point at which an additional cluster does not capture the pulse differences better, and over-fitting starts. For this work this value is always found to be 3 clusters, but in more complex problems with high dimensional parameters, could definitely be expected to increase. 

\section{Pulse interpolation}
\label{sec:pulseinterpolation}
After the originally optimized pulses have been partitioned into families by the clustering algorithm, the following step is to extend the families to the entire parameter space by means of interpolation. In this work, the parameter space will always be $\theta\in[0,\pi]$ so that we can refer to angles. For multiparameter gates, an analogous approach can be pursued.\\

\begin{figure}
    \centering
    \includegraphics[scale=0.45]{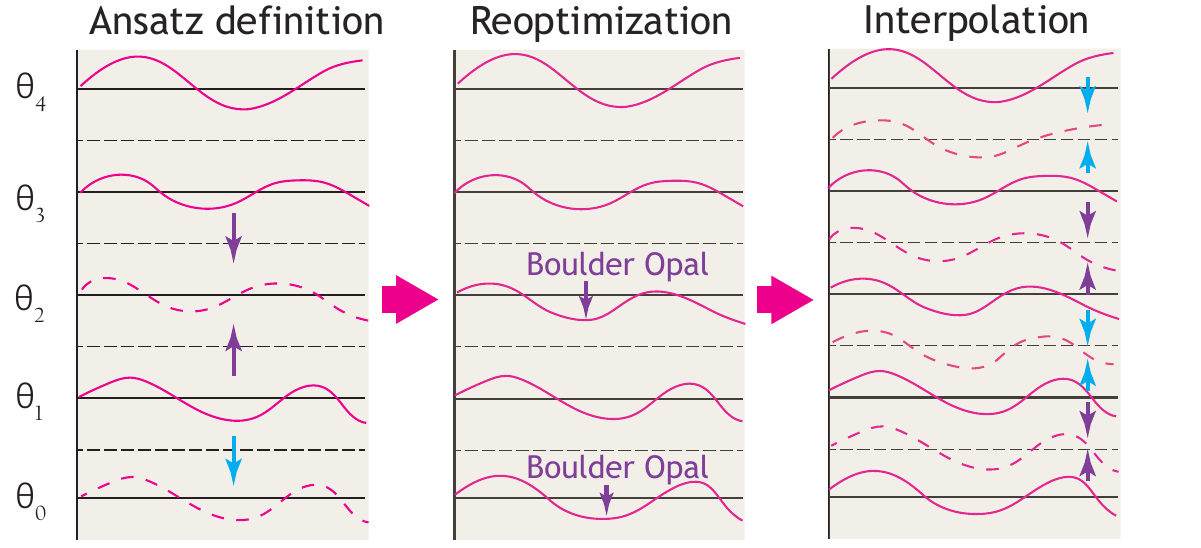}
    \caption{A family of pulses is found, defined at a subset of the predefined angles $\{\theta_i\}_i^M$, and is to be extended to all predefined angles. First, ansatzes are created by interpolation if the unassigned angle is within the range of assigned angles, or else set as the highest or lowest found pulse. These ansatzes are optimized using Boulder Opal to extend the family to all predefined angles $\theta_i$. Pulses can then be interpolated within a family in order to find high fidelity pulses for the entire parameter space.}
    \label{fig:interpolation}
\end{figure}

For each family, we want to first construct pulses on all of $\{\theta_i\}_i^M$, instead of only the subset assigned to it in the clustering. Consider one of the families, there will be a lowest and highest angle assigned to this family in the clustering. Note that a family does not necessarily include only subsequent angles, but might include gaps between disjoint regions. If an angle $\theta_i$ is not assigned to this family and is in such as gap, we take a linear interpolation as an ansatz and optimize the pulses using Boulder Opal. For fixed angles $\theta_i$ outside this range, we simply use the lowest or highest angle pulse as an ansatz and optimize. Employing these ansatzes, hopefully a pulse is found for that parameter belonging to the same family (reminiscent of Tikhonov regularization \cite{families1}). This leaves us with several extended families of pulses $Z_j$ on all predefined parameters $\{\theta_i\}^M_{i}$ (see Fig.~\ref{fig:interpolation}).\\

Using linear interpolation within a single family, new pulses for all parameters in the parameter space can be constructed, finalizing our construction of a continuous gate set for each family. Linear interpolation is computationally very efficient, especially when compared to constructing pulses from the ground up. By interpolating within a family, we expect higher fidelities than if we were to interpolate on the original set of pulses.

\textcolor{white}{.}
\\

\section{Results}
\label{sec:results}

\subsection{Distance function comparison}
\label{sec:wasserstein}
This section contrasts the effectiveness of clustering families for the Wasserstein-2 and $L_2$ distances. For arbitrarily optimized pulses, there is no established family structure to validate against. In order to straightforwardly compare the distance measures, we construct $J$ mock families of pulses $Z_j=\{z_{\theta_{j,i}}\}_i^M$. The pulses within one family are constructed in such a way that pulses $z_{\theta_i}$ and $z_{\theta_j}$ look alike for $i$ close to $j$, as they would in actual applications. The exact details of this are left for App.~\ref{app:pulseconstruction}. For each angle $\theta_i$, a corresponding pulse $z_{\text{true},i}$ from a family $Z_{\text{true},i}\in\{Z_1,...,Z_J\}$ is randomly picked for the final pulses. This mimics the pulses found by the Boulder Opal optimization procedure and accrues a set of pulses as in Fig.~\ref{fig:clustering}b, of which we have knowledge on the family structure.

On these mock pulses, the distance metrics are compared by constructing a distance matrix as in Fig.~\ref{fig:clustering}, and having the spectral clustering algorithm assign to each pulse $i$ a family $Z_{dist,i}\in\{Z_1,...Z_J\}$. The conditional probability $\mathbb{P}_{dist}$ of finding a correct match is approximated as 
\begin{equation}
    \mathbb{P}(\text{cor.}|\text{fnd.})\approx\frac{\sum_{i,j}{\mathbbm{1}[Z_{dist,i}=Z_{dist,j}]\mathbbm{1}[Z_{true,i}=Z_{true,j}]}}{\sum_{i,j}\mathbbm{1}[Z_{dist,i}=Z_{dist,j}]}.
\end{equation}
The reason for choosing this figure of merit is that when two pulses are found to be matching, they only lead to high fidelity as long as they belong to the same family. Missing a match because two pulses from the same family are clustered in different families will not lead to faulty interpolations, and is thus less important from a fidelity perspective (nevertheless, the Wasserstein-2 distance also outperforms the $L_2$ distance in terms of missed matches). Figure~\ref{fig:clustering} shows the comparison between the two distances. Wasserstein-2 can be seen to largely outperform the standard $L_2$ distance, as it is able to capture the pulse family characteristics better due to its transport based nature. In the rest of this work, the Wasserstein-2 distance is employed for clustering.

\begin{widetext2}
    \begin{minipage}[b]{\linewidth}
        \begin{figure}[H]
            \centering
            \includegraphics[scale=0.421]{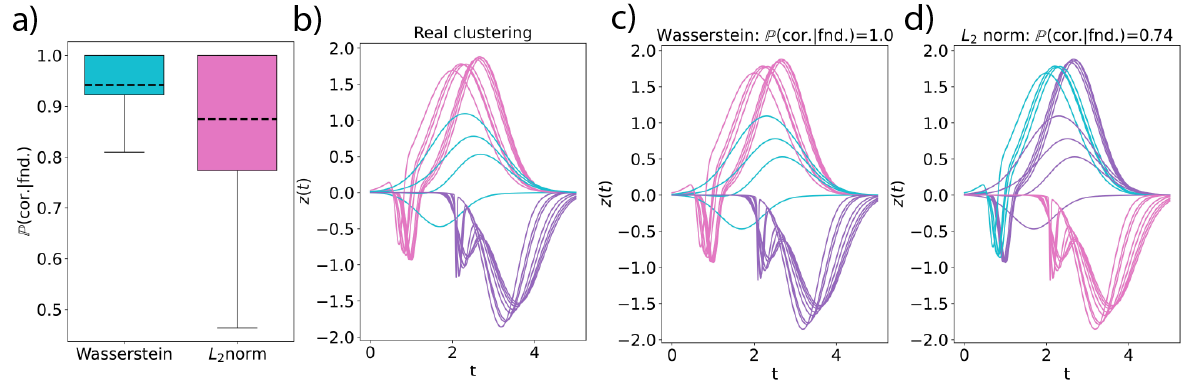}
            \caption{Comparison of the Wasserstein-2 and $L_2$ distance for pulse clustering. a) Box plots of the probability of finding a correct match $\mathbb{P}(\text{cor.}|\text{fnd.})$ for both distances, data from 2000 trials with $M=20$ angles in the interval $[0,\pi]$ randomly selected from three pulse families. b) Example of the authentic clustering of optimized pulse solutions. c) Clusters found by the Wasserstein-2 distance for the pulses from b). d) Clusters found by the $L_2$ distance for the pulses from b). }
            \label{fig:clustering}
        \end{figure}    
    \end{minipage}
\end{widetext2}

\subsection{Interpolation of gates}
\label{sec:interpolation}
To test our clustering method for the interpolation of pulses, we first optimize for $M=20$ equidistant angles $\theta_i\in[0,\pi]$ for $R_X(\theta), R_Y(\theta), R_Z(\theta)$ and $R_{ZZ}(\theta)$ for both Rydberg and cat qubits. This allows us to gather results on a universal set of gates on two vastly different systems, showing the versatility and robustness of our method.

The pulses are clustered using our spectral clustering method from Sec.~\ref{sec:clustering}, where the Elbow method indicates that for all considered applications 3 clusters is the optimal number. The families found are extended to the entire $M=20$ predefined angles $\{\theta_i\}_i^M$ using the methods from Sec.~\ref{sec:pulseinterpolation}. Lastly, for each family, we interpolate the pulses halfway the predefined angles, and determine their fidelity using~\eqref{eq:cost}.

Figure~\ref{fig:rycatfull} shows the results for the $R_Y(\theta)$ gate on cat qubits. Generally, the $R_Y$ gate is the most interesting of the three qubit gates because it requires both amplitude and phase control. In Fig.~\ref{fig:rycatfull}a, we see that when optimizing using Boulder Opal, both the original pulses and the clustered families reach high fidelities, but the interpolation on the original pulses is very bad for almost all angles. For the clustered families, on the other hand, those with good interpolations are found for the entire parameter space. This is highlighted in Fig.~\ref{fig:rycatfull}b, where the parameter space is partitioned according to where each family performs best, leading to great improvements in the  interpolation fidelities compared to the original pulses. Figures.~\ref{fig:rycatfull}c and ~\ref{fig:rycatfull}d further illustrate this  by showing that the original pulses are dissimilar for subsequent angles, resulting in low fidelities, whereas the pulses from the families in the highlighted areas correspond well and thus result in high fidelities.

Similar results can be seen in Fig.~\ref{fig:rzzrydfull} for the 2-qubit $R_{ZZ}(\theta)$ on a Rydberg system. Here we see good interpolations of the original pulses around $\theta=\pi/2$, which is also retrieved for one of the families. However, for low and high angles the interpolation fidelities become quite low for the original pulses (likely due to the presence of many local minima), but high fidelity pulses ones are constructed for the clustered families.

Figure~\ref{fig:fullresults} shows all fidelities of interpolated pulses for the problems considered. Across all parametrized gates for both qubit types, there is an increase of half up to a full order of magnitude in fidelity. Partitioning the parameter space with specific families (Clustered, pink results) results in a big advantage in both mean and variance of the fidelities compared to individual clusters (purple). Here individual clusters would be comparable to a Tikhonov regularization method as in \cite{families1}, since ansatzes for pulse optimization are used. Along the same lines, especially note the much smaller variances of the fidelities indicating that our partition clustered method yields good pulses over the entire parameter space instead of only in a specific region. The individual family method definitely results in good interpolations for certain regions of parameter space, but multiple families seem to be necessary to patch together a faithful continuous gate set over the entire parameter space. This hails especially true for the cat qubits, as seen from the fidelities in Fig.~\ref{fig:fullresults}, which is a more complex optimization problem where likely many local minima in parameter/fidelity space exist.

\begin{widetext2}
    \begin{minipage}[b]{\linewidth}
        \begin{figure}[H]
            \centering
            \includegraphics[scale=0.422]{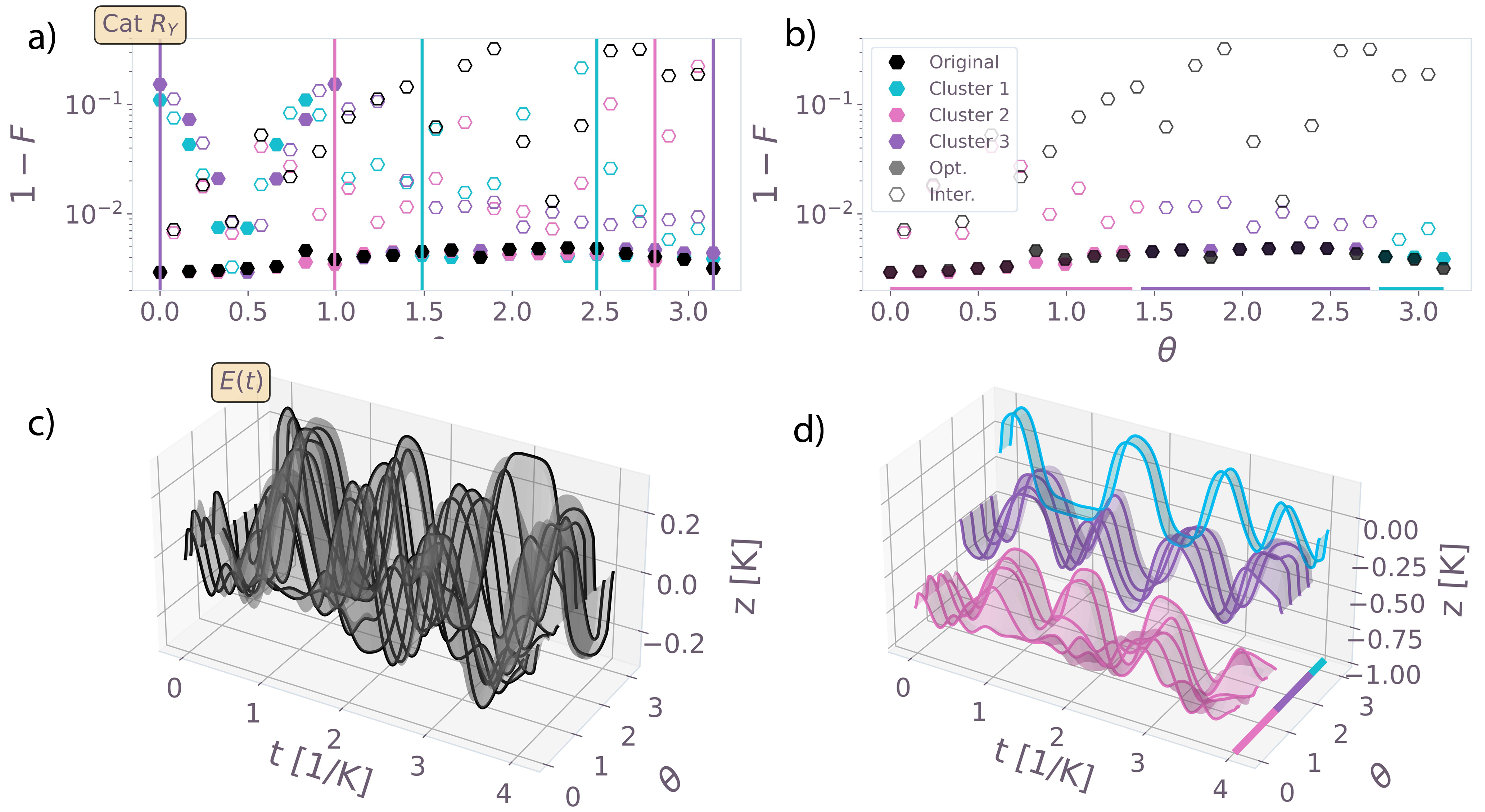}
    \caption{Optimization and interpolation results for  $R_{Y}(\theta)$ gate on a cat qubit system. a) Black, original pulses and interpolations. Colored pulses show optimized and extended clusters plus interpolations, showing matching or improved interpolation infidelities. Vertical lines are min and max angle of each cluster in the original set of pulses. b) Narrowed down results from a) where the best clusters for each regime of angles are shown, highlighting improvements in fidelity. c) Single photon drive $E(t)$ for the original set of pulses, showing a jagged interpolation landscape leading to low fidelities. Horizontal lines indicate chosen regime for each cluster in d). d) Clustered pulses for the chosen regimes, showing more regular landscape, necessary for faithful interpolations. Individual clusters are somewhat displaced on $z$-axis for visibility.}
            \label{fig:rycatfull}
        \end{figure}    
    \end{minipage}
\end{widetext2}

\begin{widetext2}
    \begin{minipage}[b]{\linewidth}
        \begin{figure}[H]
            \centering
            \includegraphics[scale=0.422]{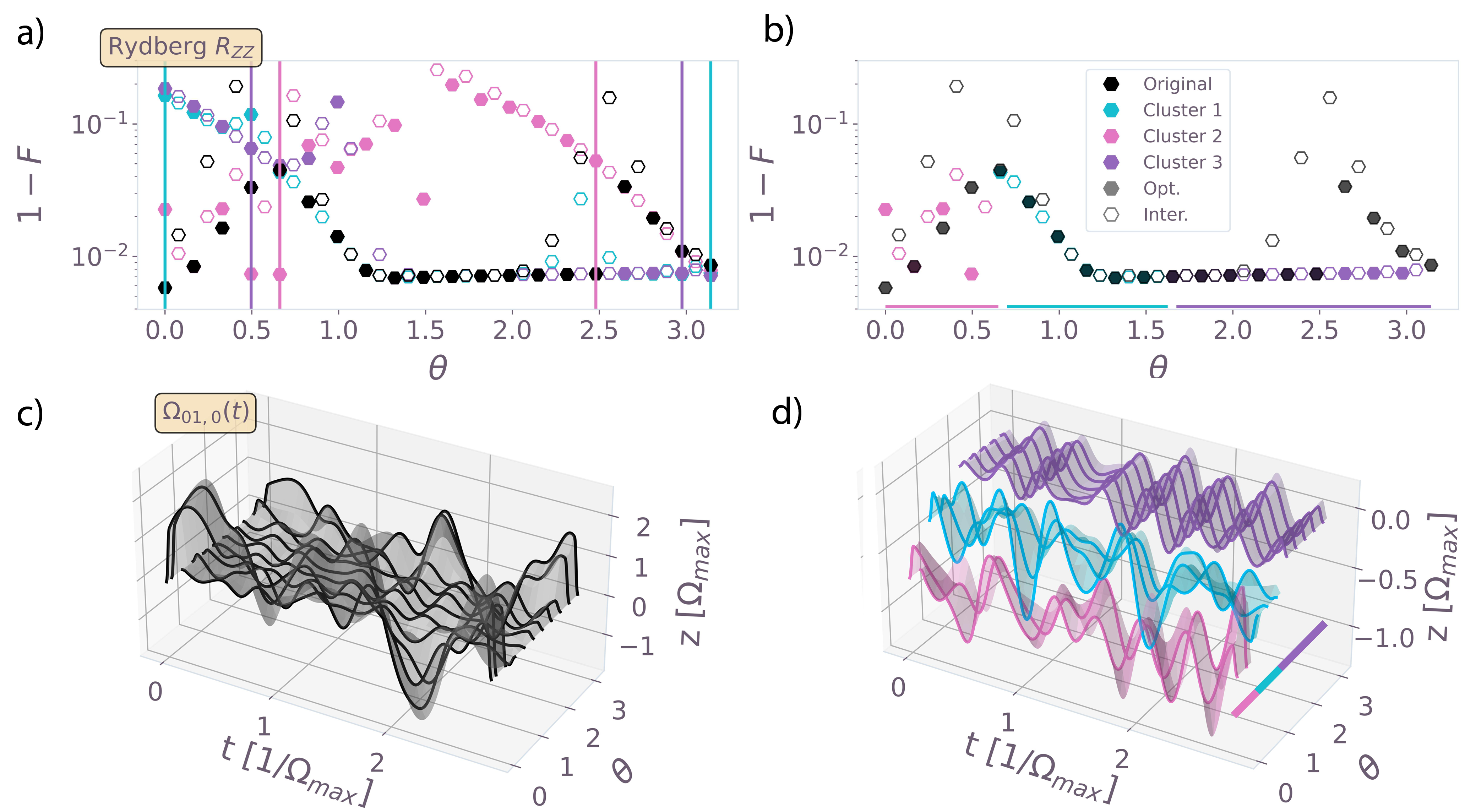}
    \caption{Optimization and interpolation results for a $R_{ZZ}(\theta)$ gate on a Rydberg neutral atom qubit system. a) Black, original pulses and interpolations, for which interpolations have low fidelities for small and large angles. Colored pulses show optimized and extended clusters plus interpolations, showing matching or improved interpolation infidelities. Vertical lines are min and max angle of each cluster in the original set of pulses. b) Narrowed down results from a) where the best clusters for each regime of angles is chosen, highlighting improvements in fidelity. c) Coupling strength of $|0\rangle\leftrightarrow|1\rangle$ transition on qubit 0 $\Omega_{01,0}(t)$ for the original set of pulses, showing a jagged interpolation landscape leading to low fidelities. Horizontal lines indicate chosen regime for each cluster in d). d) Clustered pulses for the chosen regimes, showing more regular landscape, necessary for faithful interpolations. Individual clusters are somewhat displaced on $z$-axis for visibility.}
            \label{fig:rzzrydfull}
        \end{figure}    
    \end{minipage}
\end{widetext2}

\textcolor{white}{.}
\newpage
\textcolor{white}{.}
\newpage

\begin{widetext2}
    \begin{minipage}[b]{\linewidth}
\begin{figure}[H]
\begin{centering}
    \begin{minipage}[t]{.42\textwidth}
         \hypertarget{fig:2qubitdecaya}{}
    \includegraphics[width=\columnwidth]{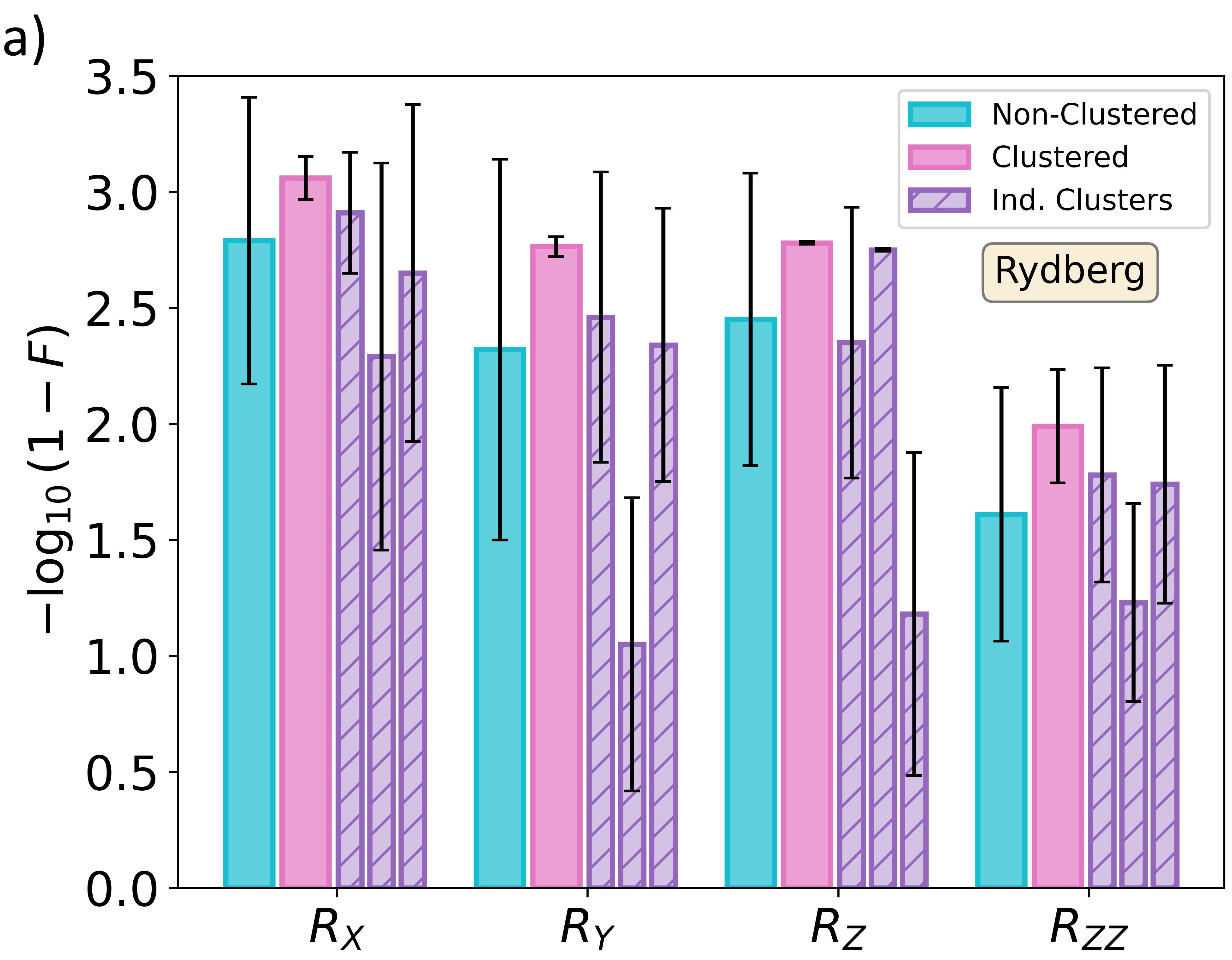}
    \end{minipage}
    \quad\quad
    \begin{minipage}[t]{.42\textwidth}
         \hypertarget{fig:2qubitdecayb}{}
         \includegraphics[width=\columnwidth]{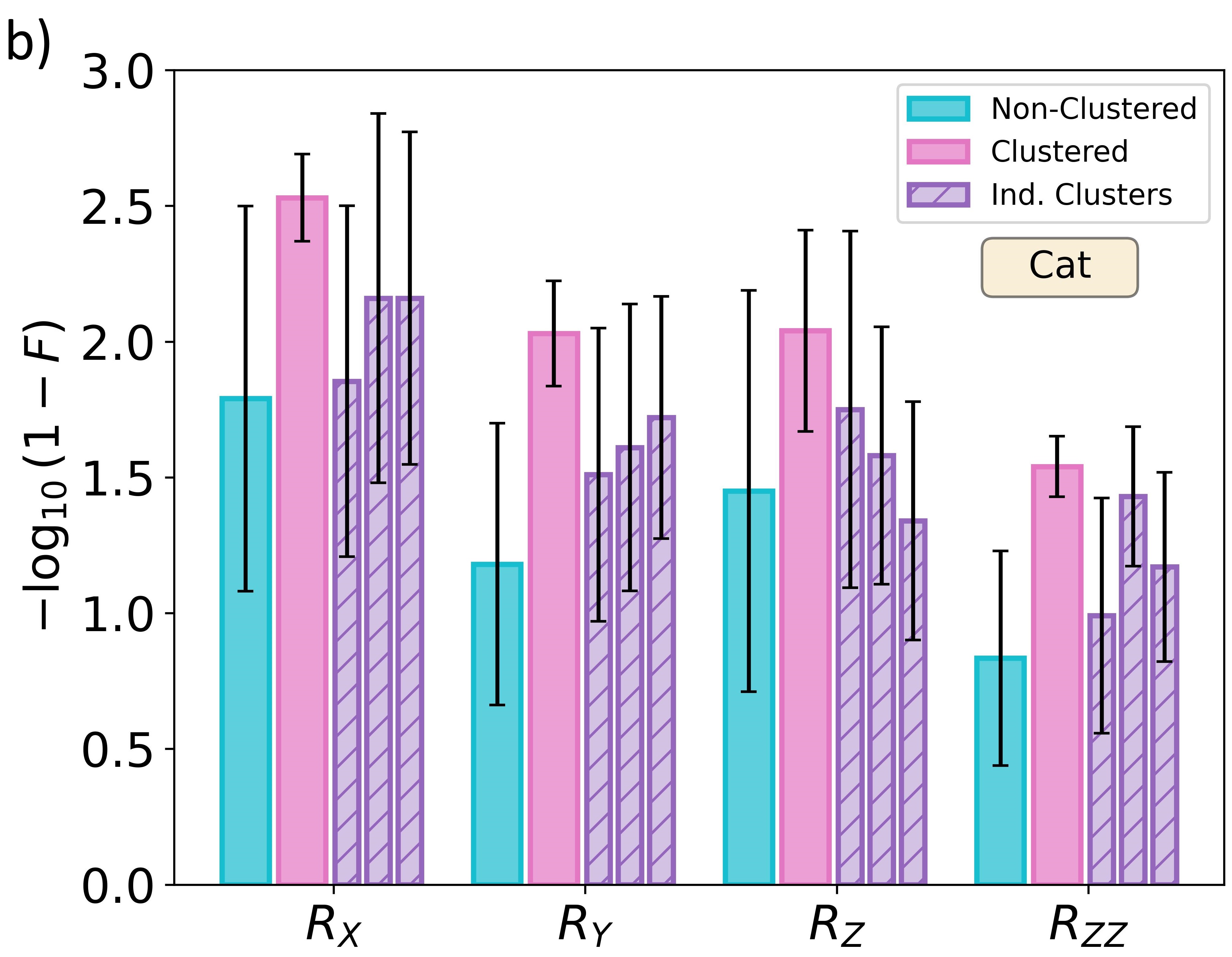}
    \end{minipage}%

\caption{Average and standard variations of infidelities of interpolations halfway between optimized angles, as in Figs.~\ref{fig:rycatfull} and~\ref{fig:rzzrydfull}. Non-clustered pulses are the original pulses (cyan), three clustered are shown (purple) as well as the fully clustered method where the parameter space is split up in regimes where the best cluster is assigned (pink). Note the higher mean fidelities as well as much lower standard variations, indicating better interpolations over the entire parameter space. a) Rydberg gates, b) Cat qubit gates.}
    \label{fig:fullresults}
    \end{centering}
\end{figure}
    \end{minipage}
\end{widetext2}

\section{Conclusion}
\label{sec:conclusion}
This work discusses and analyzes a new method for constructing continuous parameter gate sets based on the clustering of pulse families. This is an important problem as for virtually all NISQ era applications parametrized gates are required, and optimizing for each individual parameter necessary in a problem is extremely computationally expensive and slow. In our method, spectral clustering using a similarity measure based on Wasserstein-2 distances results in multiple families of pulses on which inter-family interpolation can be performed. The advantage of this method is that multiple families of pulses are found from which the best ones can be selected for a specific regime of parameter space. This method will prove to be useful for experimental realization, where one family is possibly easier to implement than another. On all gates and on both qubit systems considered, our method leads to significantly better fidelities and much more consistent interpolations than would be the case for original pulse optimization or Tikhonov based feedforward methods (individual family). In this, we highlight the great performance in cat qubits system, which shows the importance of our method in increasingly more important complex qubit schemes.\\

In future work, we hope to extend these methods to multiple parameter gates, which might be implemented in more complex VQA or QAOA type problems. Furthermore, we are curious to see the influence of directly optimizing the fidelity via the quantum channel, thus optimizing more directly for the losses, instead of optimizing the unitary evolution and post-processing the influences of the Lindbladian terms. Different parameters have different optimal gate end times, we wish to extend our methods to accommodate for varying gate times instead of one fixed optimal gate time for the largest parameter value ($\theta=\pi$ in this work). Lastly, we hypothesize that a more sophisticated way of interpolation, potentially based on Wasserstein-2 transport or on the Hamiltonian dynamics, will yield better fidelities for the interpolated pulses. 


\section*{ACKNOWLEDGEMENTS}
We thank Klaas Wijnsma, Jasper Postema, Julian Teske, Kisa Barkemeyer, Kara Maller, and Oliver Tse for fruitful discussions. This research is financially supported by the Dutch Ministry of Economic Affairs and Climate Policy (EZK), as part of the Quantum Delta NL program,  the Horizon Europe programme HORIZON-CL4-2021-DIGITAL-EMERGING-01-30 via the project 101070144 (EuRyQa), and by the Netherlands Organisation for Scientific Research (NWO) under Grant No.\ 680.92.18.05.

\section*{COMPETING INTERESTS}
The authors declare no competing interests.

\section*{DATA AVAILABILITY}
The data supporting the findings  are available from the corresponding author upon reasonable request.

\section*{CODE AVAILABILITY}
The code supporting the findings are available from the corresponding author upon reasonable request.

\newpage

\bibliographystyle{apsrev4-1}
\bibliography{Bibliography.bib}
\newpage
\onecolumngrid

\appendix
\section{Qubit systems}
\label{app:hamiltonians}
In this appendix we describe the qubit systems resulting in the system Hamiltonians $H_{\text{sys}}$, jump operators $V_k$ and control pulses $z(t)$. Furthermore, we describe the pulse optimization procedure in more detail.\\

In order to restrict the parameter space and consequently enhance the likelihood of yielding pulses in the same family, we set the pulses for qubit 1 equal to those of qubit 0 in the $R_{ZZ}$ two qubit case.  All pulses are discretized as 30 segment piecewise constant functions, which are smoothened via a Gaussian kernel and are equated to 0 at their beginning and end values, lastly being put through a low-pass filter. For each problem analyzed, the pulse end time $T$ is determined as the end time for which the problem at $\theta=\pi$ has the highest fidelity. The existence of such an optimal time follows from the reasoning that for too large gate times, decoherence effects dominate and decrease fidelity. On the other hand, if the gate time is too small, the quantum speed limit is not reached \cite{qsl} and there is no way of constructing a minimizing pulse, leading to low fidelities. The quantum speed limit likely has the highest value for $\theta=\pi$ because it least resembles the identity. To ensure we reach the quantum speed limit for all $\theta$, we choose $T$ for all problems as the optimal time for $\theta=\pi$. 

\subsection{Rydberg neutral atoms}

This section introduces basic Rydberg physics to identify what control pulses can look like for this system, as discussed in Sec.~\ref{sec:pulseoptimization}. In general, this will yield the Hamiltonian and possible Lindblad operators for the systems, as in~\eqref{eq:lindblad}. A Rydberg system is a system of individual neutral atoms trapped in laser optical tweezers, where the electronic states encode for the qubit manifold \cite{rydberg1}. For this work, we assume gg qubits \cite{rydbergqubit} such that for each qubit we have the states $|0\rangle, |1\rangle$ and $|r\rangle$, leading to find $d=3^N$ as in~\eqref{eq:cost}.

\medskip

To perform single qubit transitions on qubit $j$ between states $|a\rangle$ and $|b\rangle$, a laser interacts with the atom to realize the Hamiltonian \cite{rydbergqubit,rydberg1,rydberg2}
\begin{equation}
\begin{aligned}
    H_{j}^{ab}=\frac{\Omega_{ab,j}(t)}{2} \left(e^{i \varphi_{ab,j}(t)}|a\rangle_j\langle b|_{j}+e^{-i \varphi_{ab,j}(t)}|b\rangle_j\langle a|_{j}\right)-\Delta_{b,j}(t)|b\rangle_{j}\langle b|_j.
\end{aligned}
    \label{eq:qubitlightinteraction}
\end{equation}
Here, $\Omega_{ab,j}(t)$ denotes the coupling strength, $\varphi_{ab,j}(t)$ the phase of the laser coupled to atom $j$, and $\Delta_{b,j}(t)$ = $\omega_{ab,j}(t)-\tilde{\omega}_{ab}$ the detuning of the laser frequency $\omega_{ab,j}(t)$ from the energy level difference $\tilde{\omega}_{ab}$. In current Rydberg systems, one has exquisite control over $\Omega$ and $\Delta$, and less over $\varphi$ \cite{phase1}, thus we set $\varphi=0$. For our systems, we assume transitions $|0\rangle\leftrightarrow|1\rangle$ and $|1\rangle\leftrightarrow|r\rangle.$ This renders control pulses $z(t)\in\{\Omega_{01,j}(t),\Delta_{1,j}(t),\Omega_{1r,j}(t),\Delta_{r,j}\}$. Notice that having both coupling and detuning allows for full control on the Bloch sphere of each individual qubit, allowing for \textit{rotational control} \cite{rotational}. For this reason (and to restrict parameter space), we only optimize for $\Omega_{01}$ and $\Delta_1$ for the single qubit gates and set the other pulses equal to $z(t)=0$.  

\medskip

The Rydberg states $|r\rangle$ are high-energy excited states that have a passive `always-on' interaction, which is described by a non-controllable Hamiltonian $H_V$ \cite{rydbergqubit} as a Van der Waals interaction (VdW) \cite{vdwaals}
\begin{equation}
\begin{aligned}
    H_{V}&=\sum_{i=1}^m\sum_{j=1}^m\frac{C_6}{R_{ij}^6}|rr\rangle_{ij}\langle rr|_{ij},\\
    \label{eq:rydbergvdwinteraction}
    \end{aligned}
\end{equation}
where $R_{ij}$ is the interatomic distance and $C_6$ is an interaction coefficient. For the total Hamiltonian $H_{\text{sys}}$
\begin{equation}
    H_{\text{sys}}=\sum_j H_{j}^{01}+H_{j}^{1r}+H_V.
\end{equation}

Furthermore, the Rydberg states have a finite lifetime before decaying to $|1\rangle$. For completion, we also introduce decay from $|1\rangle$ to $|0\rangle$, which results in the jump operators
\begin{equation}
    V_{1,j}=|0\rangle_j\langle 1|_j,\quad V_{r,j}=|1\rangle_j\langle r|_j.
\end{equation}

In the optimization procedure, $\Omega_{\text{max}}=10$MHz defines the timescale, and the pulses are constrained as $|z(t)|\leq \Omega_{\text{max}}.$ Furthermore, we set $\gamma_1,\gamma_r=0.01\Omega_{\text{max}}$ which will incur fidelity losses. Lastly,  $C_6/R^6=10^3\cdot\Omega_{\text{max}}$ is taken (we only consider up to two qubits in this work). The choices for physical parameters are inspired by \cite{errorbudget} and summarized in Table~\ref{tab: optimizable pulse parameters 1}. 

\begin{table}[H]
\caption{Optimizable pulses parameters used in the optimizations. The minima and maxima values of each drive are inspired by \cite{errorbudget} The columns for each gate denote if the pulse values are real ($\mathbb{R}$), complex ($\mathbb{C}$) or kept constant (const.). When kept constant, the values are equal to 0. All units in $[\Omega_{\text{max}}]$.}
\centering
\begin{tabular}{ |P{2cm}||P{1.5cm}|P{1.5cm}|P{2.5cm}||P{1.1cm}|P{1.1cm}|P{1.1cm}|P{1.1cm}|  }
 \hline
 Pulse $z(t)$ & $\mathrm{min}[z(t)]$  & $\mathrm{max}[z(t)]$ & $z(0)=z(T)$ & $R_Z(\phi)$ & $R_X(\theta)$ & $R_Y(\varphi)$ & $R_{ZZ}(\Theta)$\\
 \hline
 $\Omega_{01}(t)$     & -1.0 & 1.0 & 0.0  & $\mathbb{R}$ & $\mathbb{R}$       & $\mathbb{R}$ & $\mathbb{R}$ \\
 $\Delta_1(t)$&  -1.0  & 1.0 & 0.0  & $\mathbb{R}$ & $\mathbb{R}$ & $\mathbb{R}$       & $\mathbb{R}$ \\
 $\Omega_{1r}(t)$     & -1.0  & 1.0 & 0.0  & const. & const.     & const. & $\mathbb{R}$ \\
 $\Delta_{r}(t)$     &  -1.0  & 1.0 & 0.0  &  const.      &  const.        & const.     & $\mathbb{R}$ \\

\hline
  \multicolumn{3}{c|}{}  & $T [1/\Omega_{\text{max}}]$  &  1.36      &  0.73        & 1.36    & 2.83 \\
\cline{4-8}
\end{tabular}
\label{tab: optimizable pulse parameters 1}
\end{table}

\subsection{Cat qubit Hamiltonians}
Our discussion of cat qubits largely follows the explanations from \cite{catqubitmain}. The cat qubit can be realized in a Kerr parametric oscillator with a two-photon pump \cite{kerr1,kerr2,kerr3}. In the rotating frame of the two-photon pump frequency $\omega_p$, the main system is described by 
\begin{equation}\label{eqn: system Hamiltonian}
\hat{H}_1=-K \hat{a}^{\dagger 2} \hat{a}^2+G_j(t)\left(\hat{a}^{\dagger 2}+\hat{a}^2\right)
\end{equation}
with $K$ the Kerr non-linearity which is constant and chosen as our timescale, $G_j$ the two-photon drive amplitude, $\hat{a}$ the annihilation operator and $\hat{a}^\dagger$ the creation operator. Rewriting the system Hamiltonian gives
\begin{equation}\label{eqn:Hamiltonian rewritten}
H_{1,j}=-K\left(\hat{a}^{\dagger 2}-\frac{G_j(t)}{K}\right)\left(\hat{a}^2-\frac{G_j(t)}{K}\right)+\frac{G_j(t)^2}{K}.
\end{equation}
With the knowledge that $\hat{a}|\text{$\pm$} \alpha\rangle = \text{$\pm$} \alpha |\text{$\pm$} \alpha\rangle$ one can see that the degenerate eigenstates of the Hamiltonian are the coherent states $|\pm\alpha\rangle$ with eigenenergy $G^2/K$. Since these eigenstates are degenerate, a linear combination 
\begin{equation}\label{eqn: cat states}
\left|C_\alpha^{+}\right\rangle=N_{\pm}(|\alpha\rangle \pm|-\alpha\rangle)
\end{equation}
of these states is also an eigenstate. The states from~\eqref{eqn: cat states} are the so-called cat states, with $N_\pm$ a normalization constant. The orthogonality of the cat states from~\eqref{eqn: cat states} allows for the following encoding
\begin{equation}\label{eqn:computational basis}
    |0\rangle = \frac{|C^+_\alpha\rangle + |C^-_\alpha\rangle}{\sqrt{2}}, \quad |1\rangle = \frac{|C^+_\alpha\rangle - |C^-_\alpha\rangle}{\sqrt{2}},
\end{equation}
as our computational basis, with $|0\rangle$ and $|1\rangle$ denoting the computational basis states. The corresponding Pauli matrices are defined through the computational basis states, e.g. $Y=i|0\rangle\langle1|-i| 1\rangle\langle0|$. The computational basis is chosen this way, foremost because single photon-loss (the dominant error mechanism in the system) is exponentially suppressed in $\alpha$. Because we can not simulate the entire Fock space, we truncate its dimension at 20 states for single qubit gates and at 16 for $R_{ZZ}$.\\

The control of these systems is described through the following Hamiltonians
\begin{equation}
    H_{2,j}=-\Delta_j(t) \hat{a}_j^{\dag} \hat{a}_j+E_j(t)(\hat{a}_je^{-i\theta_j} + \hat{a}_j^\dag e^{i\theta_j}),\quad\quad
    H_3=g(t)(\hat{a}_1\hat{a}_2^{\dag} + \hat{a}_1^{\dag}\hat{a}_2),
\end{equation}
where a detuning $\Delta_j(t)$ is created between the two-photon drive and the resonator $\Delta = \omega_r - 2 \omega_p$, with $\omega_r$ and $\omega_p$ the resonator and two-photon pump frequency respectively. $\theta_j$ the phase of the drive, which is chosen at $\theta_j=0$ (except for $R_Y$). $E_j(t)$ is the amplitude of the single-photon drive. $g(t)$ is a two-photon exchange between the two resonators. When $|g(t)|,|E_j(t)|,|\Delta_j(t)|\ll G$ the computational states are approximately kept in the computational basis. Together with single photon loss, dynamics to outside the qubit manifold is the main source of error. This leaves us with the pulses $z(t)\in\{E_j(t),\Delta_j(t),G_j(t)\}$. The total system Hamiltonian takes the form
\begin{equation}
    H_{\text{sys}}=\sum_j H_{1,j}+H_{2,j}+H_3.
\end{equation}
Single photon losses on qubit $j$ are obviously modelled by an annihilation operator $a_j$, resulting in a jump operator of the form $V_j=a_j$.

\begin{table}[H]
\caption{Optimizable pulses parameters used in the optimizations. The minima and maxima values of each drive are based on the minima and maxima found in the paper \cite{catqubitmain}. The columns for each gate denote if the pulse values are real ($\mathbb{R}$) or complex ($\mathbb{C}$). All units in $[K]$.}
\centering
\begin{tabular}{ |P{2cm}||P{1.5cm}|P{1.5cm}|P{2.5cm}||P{1.1cm}|P{1.1cm}|P{1.1cm}|P{1.1cm}|  }
 \hline
 Pulse $z(t)$ & $\mathrm{min}[z(t)]$  & $\mathrm{max}[z(t)]$  & $z(0)=z(T)$ & $R_Z(\phi)$ & $R_X(\theta)$ & $R_Y(\varphi)$ & $R_{ZZ}(\Theta)$\\
 \hline
 $E(t)$     & -0.308 & 0.308 & 0.000  & $\mathbb{R}$ & $\mathbb{R}$       & $\mathbb{C}$ & $\mathbb{R}$ \\
 $\Delta(t)$&  0.000 & 4.000 & 0.000  & $\mathbb{R}$ & $\mathbb{R}$ & $\mathbb{R}$       & $\mathbb{R}$ \\
 $G(t)$     & -4.000 & 4.000 & 4.000  & $\mathbb{R}$ & $\mathbb{R}$       & $\mathbb{R}$ & $\mathbb{R}$ \\
 $g(t)$     &  0.000 & 0.154 & 0.000  &  n.a.        &  n.a.        & n.a.         & $\mathbb{R}$ \\

 \hline
  \multicolumn{3}{c|}{}  & $T [1/K]$  &  2.06     &  2.06        & 4.03    & 3.00 \\
\cline{4-8}

\end{tabular}
\label{tab: optimizable pulse parameters 2}
\end{table}

\section{Wasserstein-2 norm}
\label{app:Wasserstein}
The Wasserstein-2 distance $W_2(z_1,z_2)$ calculates the cost of moving the pulse $z_1$ onto $z_2$, and is more closely related to our intuitive understanding of similarity between pulses than the $L_2$ distance \cite{santambrogio}. To understand this distance, we first define a distribution on a pulse $\mu[z]$ as

\begin{equation}
    \mu[z](A)=\frac{1}{c[z]}\int_{(t,y)\in A} \delta(y=z[t])d(t,y),\quad A\subset\mathbb{R}^2.
\end{equation}
Thus, the measure is over $\mathbb{R}^2$ and only puts weight on the points of the pulse $(t,z(t))$. Here $c[z]$ is a normalization constant s.t. $\mu[z](\mathbb{R}^2)=1$ and is simply equal to the length of the curve, e.g. 
\begin{equation}
    c[z]=\int_{0}^T \sqrt{1+(\partial_tz)^2}dt.
\end{equation}
We now want to transport the distribution $\mu[z_1]$ to the distribution $\mu[z_2]$, and for that purpose define a coupling $\gamma\in\Pi(\mu[z_1],\mu[z_2])$, which is a joint probability measure on $\mathbb{R}^2\times\mathbb{R}^2$ whose marginals are $\mu[z_1]$ and $\mu[z_2]$ \cite{villani} (see Fig.~\ref{fig:wasserstein}), e.g.
\begin{equation}
    \int_{\mathbb{R}^2}\gamma(x,y)dy=\mu[z_1](x),\quad \int_{\mathbb{R}^2}\gamma(x,y)dx=\mu[z_2](y).
\end{equation}
Here, $\Pi(\mu[z_1],\nu[z_1])$ is the set of all transport couplings between  distributions $\mu[z_1]$ and $\mu[z_2]$. The Wasserstein-2 distance $W_2(z_1,z_2)$ is then defined as the minimal coupling cost
\begin{equation}
\label{eq:wmetric}
    W_2(z_1, z_2)^2:=\min_{\gamma\in \Pi(\mu[z_1],\mu[z_2])} \int_{\mathbb{R}^2\times \mathbb{R}^2}|x-y|^2 \mathrm{~d} \gamma(x,y).
\end{equation}

\section{Pulse Construction}
\label{app:pulseconstruction}
In order to construct a mock family of pulses, we define for $\theta_0=0$ a sum of $G$ Gaussians where each Gaussian $g$ has a random amplitude $a_{0,g,1}$, center $a_{0,g,2}$ and variance $a_{0,g,3}$ chosen according to some uniform random distributions. For subsequent $\theta_{i+1}$, we then have these coefficients evolve according to
\begin{equation}
a_{i+1,g,k}=a_{i,g,k}+\mu_{g,k}\Delta\theta+\sigma_{g,k}\mathcal{N}\sqrt{\Delta\theta},
\end{equation}
where $\mu_{g,k}$ and $\sigma_{g,k}$ are predefined drifts and variances, respectively. $\mathcal{N}$ is a normally distributed random variable and $\Delta \theta$ is the step size. These kinds of pulses mimic the pulse optimization solutions we get when constructing pulses for a continuously parametrized gate $U(\theta)$. When the step size, drift and variances are high, subsequent pulses look less alike and the clustering problem becomes harder to solve, leading to lesser results for both Wasserstein-2 and $L_2$ distances. Nevertheless, in all tried out configurations we saw Wasserstein-2 outperform $L_2$, especially when subsequent pulses are more alike, which can always be achieved by lowering $\Delta \theta$.

\end{document}